\documentclass[usenatbib]{emulateapj}
\usepackage{xcolor}
\usepackage{amsmath}
\usepackage{graphicx}
\usepackage[titletoc]{appendix}
\newcommand{\degree}{$^{\circ}$}
\usepackage[unicode=true,pdfusetitle,bookmarks=true,bookmarksnumbered=false,bookmarksopen=true,breaklinks=true,pdfborder={0 0 0},backref=section,colorlinks=true]{hyperref}
\hypersetup{linkcolor=blue,citecolor=blue}

\makeatother
\shorttitle{Velocity Gradients as a Tracer for Magnetic Fields}
\shortauthors{Gonz\'alez-Casanova \& Lazarian}

\begin{document}
\title{Velocity Gradients as a Tracer for Magnetic Fields}
\author{Diego F. Gonz\'alez-Casanova and A. Lazarian}
\email{casanova@astro.wisc.edu}
\affil{Astronomy Department, University of Wisconsin-Madison, 475 North Charter Street, Madison, WI 53706-1582, USA}
\begin{abstract} 

Strong Alfv{\'e}nic turbulence develops eddy-like motions perpendicular to the local direction of magnetic fields.  This local alignment induces velocity gradients perpendicular to the local direction of the magnetic field. We use this fact to propose a new technique of studying the direction of magnetic fields from observations, the Velocity Gradient Technique (VGT). We test our idea by employing  the synthetic observations obtained via 3D MHD numerical simulations for different sonic and Alfv{\'e}n Mach numbers. We calculate the velocity gradient, $\mathbf{\Omega}$, using the velocity centroids. We find that $\mathbf{\Omega}$ traces the projected magnetic field best for the synthetic maps obtained with sub-Alfv{\'e}nic simulations providing good point-wise correspondence between the magnetic field direction and that of $\mathbf{\Omega}$. The reported alignment is much better than the alignment between the density gradients and the magnetic field and we demonstrated that it can be used to find the magnetic field strength using the  analog of Chandrasekhar-Fermi method. This new technique does not require dust polarimetry and  our study opens a new way of studying magnetic fields using spectroscopic data.

\end{abstract}

\maketitle
\section{Introduction} 

It is well established that the interstellar medium (ISM) is turbulent, affecting the dynamics of different astrophysical phenomena such as star formation, cosmic ray acceleration, galaxy evolution, and feedback \citep{ferriere2001, elmegreen2004, deavillez2005, mckee2007, falgarone2008}. Widely-used evidence of the turbulence in the ISM is seen in the so-called Big Power Law in the Sky \citep{armstrong1995, chepurnov2010} that reflects the Kolmogorov spectrum of electron density fluctuations, in nonthermal Doppler broadening \citep[see][]{draine2011}, in the power-law scalings of the fluctuations measured in Position-Position Velocity (PPV) space \citep[see][]{lazarian2000, stanimirovic2001, chepurnov2010b, chepurnov2015, padoan2009} \citep[see][for a review]{lazarian2009}, and in velocity centroids \citep[see][]{miesch1999, miville2003}.

Study of direction and measurements of intensity of magnetic fields present a challenging problem.  Polarization arising from grains aligned with longer axes perpendicular to a magnetic field allows for the estimation of the direction of the magnetic field, but the existing uncertainties in grain alignment theory and the failure of grain alignment at large optical depths limit the ability of the technique to trace magnetic fields without ambiguities \citep[see][for a review]{lazarian2007b, lazarian2015d}. This also affects the Chandrasekhar-Fermi (C-F) technique that is being used to find the intensity of magnetic fields when the variations of polarization direction and velocity dispersion are both known. 

Other techniques that are used to study magnetic fields in molecular clouds, e.g.  based on the Zeeman effect or Goldreich-Kylafis effect have their own limitations \citep[see][]{crutcher2012}.  Faraday rotation is able measure the intensity of the field along the line of sight in ionized media \citep{padoan2001, ostriker2003, crutcher2012}. The atomic/ionic  alignment technique provides a promising approach for studies of magnetic field, but the ability of the technique to trace magnetic field has not been demonstrated with the observational data yet \citep[see][]{yan2015}.

Therefore there is intense interest in developing new techniques. For instance, the anisotropy analysis of statistical properties of variations of the observable Doppler shifted lines \citep{lazarian2002, esquivel2005, heyer2004, esquivel2009, burkhart2015} proved to be a promising way of obtaining the mean magnetic field as well as the media magnetization \citep[see also][]{esquivel2015, kandel2016, kandel2016a}. The difference of the velocities of ions and neutrals due to the difference of their damping in MHD turbulence has been suggested as another way of studying magnetic fields \citep{li2008}.The quantitative treatment of the technique shows, however, that it has its limitations\citep[see][]{xu2015}. New approaches for studying magnetic fields using synchrotron intensity and polarization have been suggested in \citet{lazarian2012a, lazarian2016} All in all, the search of new ways of studying magnetic fields is characterized by intensive research with all techniques having their shortcomings and limitations. 

The \citet[][C-F]{chandrasekhar1953} method can determine magnetic field strengths by estimating the variations of the directions of  the magnetic field lines, assuming Alfv{\'e}nic motions \citep{ostriker2001, houde2009, novak2009, falceta2008}. The C-F method in its simplest incarnation estimates the magnetic field strength as: 
\begin{equation}
B \approx \sqrt{4 \pi \rho} \; \delta v / \delta\phi \;, 
\label{eq:cf}
\end{equation}
where $\rho$ is the density, $\delta v$ is the velocity dispersion, and $\delta \phi$ is the polarization angle dispersion. An improved C-F technique  in \citep{falceta2008} was shown to be able to determine the intensity of the magnetic field with less than a $20\%$ uncertainty.  In the case when turbulence is injected on small scales \citet{cho2016} provides another modification of  C-F method.

In a strongly magnetized turbulent medium, the turbulent eddies rotate preferentially about the local direction of the magnetic field \citep{goldreich1995, lazarian1999, cho2000, cho2002a}. Thus the analysis of their rotation can determine the direction of the magnetic field. This is the basis for the new idea of studying magnetic field that we advocate in this paper. 

In this {\it paper} we introduce a new technique, the Velocity Gradient Technique (VGT) and demonstrate its ability to trace the direction of the magnetic field. For our studies we use synthetic maps obtained with 3D MHD simulations and use gradients of the first order velocity centroids\footnote{Velocity centroids can be of different orders depending on the power of velocity that enters the calculation of the centroid.}. To test the ability of the VGS to represent magnetic field we rotate these gradients 90 degrees and compare their direction with the direction of the magnetic field averaged along the line of sight.

In what follows, in Section~\ref{sec:theo}, we explore the theoretical approach to the use of velocity gradients; in Section~\ref{sec:num}, we describe the numerical code and setup for the simulations; in Section~\ref{sec:results1}, we present our method of the VGT and the alignment with the magnetic field; in Section~\ref{sec:properties}, we present observational properties of the velocity gradients; \ref{sec:c-f}, we present the use of the C-F method on the VGT; in Section~\ref{sec:discussion}, we discuss our technique; and in Section~\ref{sec:conclusion}, we give our conclusions.
\section{Theoretical Considerations}\label{sec:theo}

The VGT is based on the modern theory of MHD turbulence \citep[see][ for a review]{brandenburg2013}. In strong MHD turbulence,  the Alfv{\'e}nic turbulence motions type are eddy-like \citet{goldreich1995}, henceforth GS95).  However, unlike the hydrodynamical case, the eddy motions have a preferential direction of motion set by the magnetic field direction. The eddies mix magnetic fields and matter perpendicular to the local direction of magnetic field.  The local system of reference, which was not a part of the original GS95 picture, currently is one of the major pillars of the present-day understanding of MHD turbulence \citep{lazarian1999, cho2000, maron2001, cho2002c}. It is in this local frame, that the relation between parallel and perpendicular scales of the eddies is set by a ``critical balance":
\begin{equation} 
l_{\parallel}^{-1} V_A \approx l_{\perp}^{-1}  u_l \;,
\label{eq:cb}
\end{equation}
with $V_A$ the Alfv{\'e}n speed, $ u_l$ the eddy velocity, and $l_{\parallel}$ and $l_{\perp}$ the eddy scales parallel and perpendicular to the local direction of the magnetic field. The  Alfv\'{e}n speed $V_A$ is: $\langle |\mathbf{B}| \rangle/\sqrt{\langle \rho \rangle}$, where $\langle \cdot\rangle$ is the average over the entire data set. This critical balance determines the eddy size by the distance an Alfv{\'e}nic perturbation can propagate during an eddy turnover \citep[for a review see][]{lazarian2012}.  For the sub-Alfv{\'e}nic regime the eddy velocity and scales can be written in terms of the injection velocity ($V_L$) as \citep{lazarian1999}:
\begin{eqnarray}
l_\parallel \approx L \Big(\frac{l_\perp}{L}\Big)^{2/3} M_A^{-4/3} \;, \hfill \nonumber \\
u_l \approx V_L \Big(\frac{l_\perp}{L}\Big)^{1/3} M_A^{1/3} \;, \hfill \nonumber\\
\label{eq:eddy-velocity}
\end{eqnarray}
where $\mathit{M_A} = \langle |\mathbf{v}|/V_A \rangle$ is the Alfv\'{e}nic Mach number, $|\mathbf{v}|$ is the local magnitude of the velocity field and $L$ is the injection scale with injection velocity $V_L$. These intrinsic properties of the eddies imprinted by the Alfv{\'e}nic turbulence imply not only the condition of a preferential direction along the local magnetic field, but also that the eddy velocity depends the size of the eddy.  

The effects of the anisotropy of the velocity fluctuations on the turbulent medium have been described by analyzing intensity anisotropies of spectral line cube velocity channels \citep{lazarian2002, burkhart2014, esquivel2015, kandel2016a}, correlations of velocity centroids \citep{esquivel2005, federrath2010, kandel2016},  the bispectrum \citep{burkhart2009}, and higher order statistical moments \citep{kowal2007}, as well as using Principal Component Analysis (PCA) \citep{heyer2008}.  All these techniques, however, require that the statistical samples that limit their ability to trace magnetic fields over sufficiently small patches of the sky.

It is also important to mention, that the local system of reference cannot be studied in observations where the averaging along the line of sight is performed. The projection effects inevitably mask the actual direction of the magnetic field within individual eddies along the line of sight. As a result, the scale-dependent anisotropy predicted in the GS95 model is not valid for the observer measuring parallel and perpendicular scales of projected and averaged (along the line of sight) eddies. The anisotropy of eddies becomes scale-independent and the degree of anisotropy gets determined by the anisotropy of the largest eddies for which projections are mapped \citep{cho2002, esquivel2005}. The projections of eddies for sub-Alfv{\'e}nic turbulence is aligned along the magnetic field. 

The elongated eddies have the largest velocity gradient perpendicular to the their longest axes. Thus we expect the direction of the maximum velocity gradient to be perpendicular to the local magnetic field. Thus, the velocity gradients can trace the directions of the local magnetic field, while one can expect that the observed gradients of the Doppler shifted spectral lines will trace the plane of sky variations of magnetic fields. The measurements do not require determining the correlations and therefore can be made more local compared to the anisotropies of correlations that we have discussed above. 
\section{MHD Simulations}\label{sec:num}

We used two MHD codes to simulate the data, AMUN for the sub-Alfv{\'e}nic regime \citep{kowal2007, kowal2009} and a code developed by  \citet{cho2002} for the super-Alfv{\'e}nic regime . Both codes solve the ideal MHD Equations with periodic boundary conditions, 


\begin{eqnarray}
\frac{\partial \rho}{\partial t} + \mathbf{\nabla} \cdot (\rho \mathbf{v}) = 0 \;, \hfill \nonumber \\
\frac{ \partial \rho \mathbf{v}}{\partial t} + \mathbf{\nabla} \cdot
\Big[ \rho\mathbf{v}\mathbf{v} + \big( p + \frac{B^2}{8\pi} \big)
\mathbf{I}-\frac{\mathbf{B}\mathbf{B}}{4\pi} \Big] =  \mathbf{f} \;, \hfill \nonumber \\
\frac{\partial \mathbf{B}}{\partial t} = \mathbf{\nabla} \times
(\mathbf{v} \times \mathbf{B}) \;, \hfill \nonumber \\
p = c^2_s \rho \hfill \;, \nonumber\\
 \mathbf{\nabla} \cdot \mathbf{B} = 0 \hfill \;, \nonumber\\
\end{eqnarray} 
were $\rho$ is the density, $p$ is the pressure, $c_s$ is the isothermal sound speed, $\mathbf{B}$ is the magnetic field, $\mathbf{v}$ is the velocity and $\mathbf{f}$ is the external force (in this case the turbulence injection force). 

The turbulence is driven solenoidally in Fourier space at a scale 2.5 times smaller than the simulation box size, i.e. $2.5~l_{inj} = l_{box}$. This scale defines the injection scale in our modes in Fourier space, and therefore the largest scale of the eddies. Density structures in turbulence can be associated with the effects of slow and fast modes \citep{beresnyak2005, kowal2007}. The code units of length are defined in terms of the box size ($L$), and the time as the eddy turnover time ($L/ \delta v$). For this simulation the velocity and density fields are set to unity in code units while  the pressure was changed to get the different sonic Mach numbers. The sonic Mach number is defined as: $\mathit{M_S} = \langle |\mathbf{v}|/c_s \rangle$.   The simulations are isothermal scale-free, so they can be scaled for any parameters of the observed media studied \citep[see][]{burkhart2009} provided that the cooling in the media is efficient to keep it isothermal. The properties of different phases of the interstellar medium can be found at \citet{draine1998}.   To reach saturation and stability, the simulations run for 5-7 turnover times of the largest eddy.  

The magnetic field has a uniform component ($\mathbf{B_o}$) and a fluctuating field ($\mathbf{\delta b}$), i.e. ($\mathbf{B} = \mathbf{B_o}+\mathbf{\delta b}$).  Initially $\mathbf{\delta b}=0$ and for all times $\langle \mathbf{\delta b} \rangle = 0$.  The fluctuating field is produced due to the turbulence injection.   The mean field is in the `$x$' direction. The database consists of 10 numerical simulations with a resolution of  $512^3$ and $256^3$ for the super-Alfv{\'e}nic and $192^3$ and $256^3$ for the sub-Alfv{\'e}nic regime, where the mean magnetic field has values of 10, 1, and 0.1 in the units of the driving velocity (see Table \ref{table:models}). Therefore magnetic field B=10 corresponds to the Alfv{\'e}n Mach number $M_A=V_L/V_A=0.1$, where $V_L$ is the velocity at the scale $L$ of turbulent injection, $V_A$ is the Alfv{\'e}n velocity, while the trans-Alfv{\'e}nic turbulence corresponds to $B=1$. The supersonic simulations have a resolution of $512^3$ and $256^3$ .  For most of the analysis we use the first 3 models.

\section{Alignment of velocity gradients and magnetic field} \label{sec:results1}

\subsection{Velocity Centroids}

Observational information, such as velocity, density, intensity and magnetic field, correspond to the projected information of the 3D medium. The projection is done along the line of sight (LOS) generating a 2D  the plane of sky field. 

A common technique used to study Doppler-shifted spectral lines is based on the analysis of velocity centroids. The most popular of them are the first moments of spectral line \citep[see][]{munch1958, kleiner1985,  odell1987, miesch1999}. Velocity centroids were also suggested as a means of measuring anisotropy of turbulence using velocity correlations \citep{esquivel2005, esquivel2009, burkhart2014}. A theoretical elaboration of the latter technique using the analytical description of PPV \citep{lazarian2000, lazarian2004, lazarian2008b} was obtained in \citet{kandel2016}. In the VGT we do not use correlations of centroids, but their gradients.

Velocity centroids $ C(\mathbf{x})$ and $S(\mathbf{x})$ (normalized and un-normalized respectively),\footnote{The traditionally-used centroids are normalized, but the study in \citet{esquivel2005} showed that for practical purposes, the normalization does not give much, but significantly complicates the statistical study. Thus we introduced un-normalized centroids which were used for many subsequent studies \citep[see][]{kandel2016}.} give information on the velocity field of the medium.  Apart from velocity centroids we use intensity of total emission. For our model we assume that the intensity, $I(\mathbf{x})$, is proportional to the column density, just like the case of optically thin HI emission: 
\begin{eqnarray}
C(\mathbf{x}) =\frac{ \int v_z(\mathbf{x},z) \rho(\mathbf{x},z) dz }{\int \rho(\mathbf{x},z) dz}\;, \hfill \nonumber \\
S(\mathbf{x}) = \int v_z(\mathbf{x},z) \rho(\mathbf{x},z) dz\;, \hfill \nonumber \\
I(\mathbf{x}) = \int \rho(\mathbf{x},z) dz\;, \hfill \nonumber \\
\label{eq:centroid}
\end{eqnarray}
where $\rho$ is the density, $v_z$ is the LOS component of the velocity, $\mathbf{x}$  is the position of the plane of the sky, and `$z$' is the position along the LOS.  For velocity centroids of higher order $v_z$ can enter with different power. For instance, quadratic centroids can have $v_z^2$, cubic $v_z^3$, etc. The use of higher moments increases velocity contribution to the measure, but it also can enhance the noise in the signal. In this work we only demonstrate the technique's abilities using the first order centroids. Centroids or arbitrary moments can be contracted from spectral line observations.

Using Equation \ref{eq:centroid} we construct three 2D maps with the mean magnetic field, $\mathbf{B_o}$, perpendicular to the LOS.  One map for the intensity (density) and two for the centroids (velocity).  While velocities directly trace turbulence, our earlier studies shows that density is a much more distorted tracer, especially at high Mach numbers \citep{beresnyak2005, kowal2007}.  The differences between the velocity and density gradient directions is a very important measure which relevance will be studied elsewhere. Combining the two one can not only get the information on the Mach number from the difference of directions of gradients, but provide a much better idea at which regions the velocity tracing may be distorted by shocks and other motions not related to MHD turbulence.

We compare the directions of projected velocity centroid gradients with the direction of projected magnetic fields. The magnetic fields were also projected along the LOS by:
\begin{equation}
\mathbf{B}_x(\mathbf{x}) =  \int \mathbf{B}_x(\mathbf{x},z) dz / \Delta z \hfill \;,
\end{equation}
\begin{equation}
\mathbf{B}_y(\mathbf{x}) =  \int \mathbf{B}_y(\mathbf{x},z) dz  / \Delta z   \hfill \;,
\end{equation}  
where $\mathbf{B}_x$ and $\mathbf{B}_y$ are the components of the magnetic field perpendicular to the LOS, and $\Delta z$ is the length of the LOS.  The map of the projected magnetic field is shown in the first row of Figure \ref{fig:angles}.

\begin{table}
\centering
\caption{Simulation parameters} 
\begin{tabular}{r*{9}{c}}
Model &  $\mathbf{B_o}$  & $\mathit{M_A}$ & $\mathit{M_S}$ & Resolution\\
\hline
\hline
1 & 10    &  0.1   &  0.7 & $192^3$ \\
2 & 1      &  0.7   &  0.7 & $256^3$\\
3 & 0.1   &  2.7   &  0.7 & $512^3$\\
4 & 1      &  0.7   &  1    & $512^3$\\
5 & 1      &  0.7   &  1.5    & $256^3$\\
6 & 1      &  0.7   &  3    & $512^3$\\
7 & 1      &  0.7   &  4.5    & $512^3$\\
8 & 1      &  0.7   &  5.5    & $256^3$\\
9 & 1      &  0.7   &  7    & $512^3$\\
10 & 0.1   &  2.7   &  2    & $512^3$\\
\hline
\end{tabular}
\label{table:models}
\end{table}

\subsection{The Velocity Gradient Technique}\label{subsec:vgt}

The velocity gradient information is available in observations using the velocity centroids which are the measures integrated along the line of sight.  Hence, the gradients of velocity centroids  are affected by the projection effects. The maximum gradient, that we use for the analysis $\nabla$, is defined as:
\begin{equation}
\nabla^U(\mathbf{x}) =  \rm{max}\bigg\{ \frac{|U(\mathbf{x})- U(\mathbf{x}+\mathbf{x'})|}{|\mathbf{x'}|} \bigg\} \;,
\end{equation}
where $U(\mathbf{x})$ is the projected information ($C(\mathbf{x})$ or $S(\mathbf{x})$), and $\mathbf{x'}$ is defined in a circular punctured neighborhood around the point $\mathbf{x}$ of radius $r$. For our calculations we use $r = 10$ cells.  To estimate the lag effects on the velocity gradient we modify the lag for the estimation of the velocity gradient, ranging form 2 to 64, in powers of $2^n$.  The results are independent of the lag for lags larger than $\sim8$.  Below this size, the circular neighborhood presents the effects of a square grid, slightly increasing the error on the measurement.   In other words, a larger lag results in greater precision in the determined direction (angle) of the maximum gradient.  For this pilot study we do not use an interpolation. Therefore,  in the case of $r=1$ there are only four angle options, while for $r=10$ there are over 50. {\it In practice, the calculation of the gradients should be performed on the scales larger that the scale at which numerical effects are getting important. For our simulations this scale corresponds to $r\sim8$).} 

\begin{figure}[h]
\centering\includegraphics[width=\linewidth,clip=true]{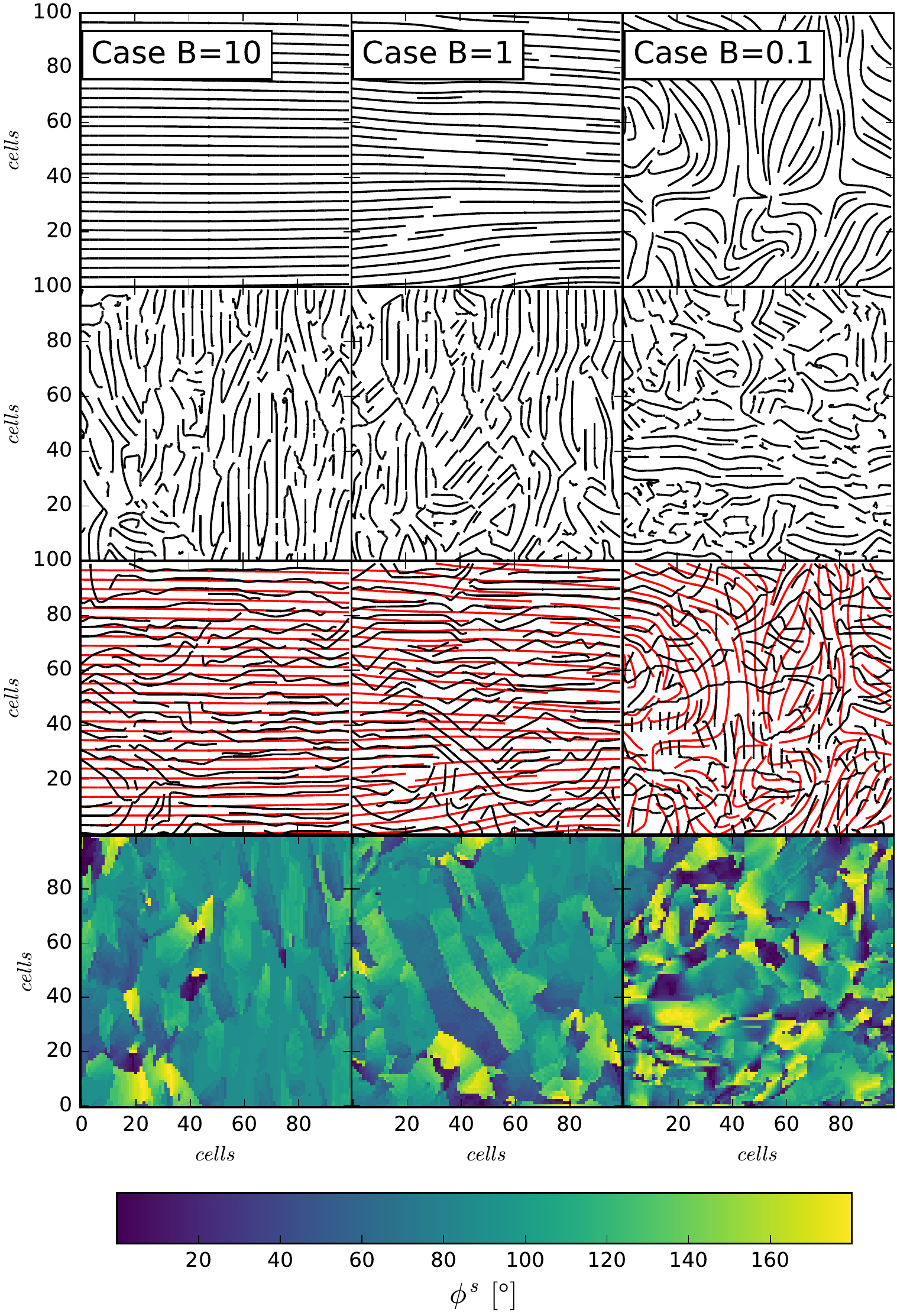}
\caption{{\it 1$^{st}$ row}: projected magnetic field, $\mathbf{B}$. {\it 2$^{nd}$ row}: gradient field for the un-normalized centroids, $\mathbf{\Omega}^S$. {\it 3$^{rd}$ row}: gradient field for the un-normalized centroids rotated 90\degree~in {\it black}, and the $\mathbf{B}$ in {\it red} for comparacing, and {\it 4$^{th}$ row}: angle between the two vector fields, $\phi^S$.  From left to right the different models decrease the magnetic field strength. All plots correspond to a subregion of $100^2$ cells of the simulation. The Figure presents how the technique of the velocity gradient looks on a Subsection of the simulation.}
\label{fig:angles}
\end{figure}

The velocity gradient field, $\mathbf{\Omega}(\mathbf{x})$, that marks the direction of the maximum gradient is defined as:
\begin{equation}
\mathbf{\Omega}^U(\mathbf{x}) =  \mathbf{x'} \;.
\end{equation}
$\mathbf{\Omega}$ is therefore constructed such that in Alfv{\'e}nic turbulence its direction is preferentially perpendicular to the local magnetic field. Figure \ref{fig:angles} (third row) shows the projected magnetic field and the rotated 90\degree velocity gradient vector. The rotation accounts for the fact, as we discussed above, that the velocity gradients tend to be perpendicular to the magnetic field. We observe a fair alignment of the magnetic field and the rotated velocity gradients obtained with velocity centroids (see Figure \ref{fig:angles}).

We treat the velocity gradient field in the way analogous to the polarization from aligned grains \citep[see][]{lazarian2007b}, i.e. we assume 180 degree ambiguity in the direction of gradients as, similar to dust grains, gradients are not sensitive to the direction of magnetic field, but give the direction perpendicular to it. To compare the alignment between the two vector fields, namely, $\mathbf{\Omega}$ and $\mathbf{B}$, one can use the angle between the two vectors, $\phi$.  $\phi$ measures the full angle span (0\degree -180\degree), giving information on the actual magnetic field vector (including the direction). 

Figure \ref{fig:angles} shows the magnetic field ($\mathbf{B}$), gradient field for the un-normalized centroid ($\mathbf{\Omega}^S$), gradient field for the un-normalized centroid rotated 90\degree to match that of the magnetic field, and the angle between the two fields for the three magnetic field intensities ($\phi^S$). Within our exploratory study, we observe that the correlation of magnetic field direction and the gradient direction is best for high degrees of media magnetization, i.e. corresponding to $B=10$ and gets worse for super-Alfv{\'e}nic turbulence with $B=0.1$. The magnetization that was used in our earlier papers \citep[see][]{esquivel2005, burkhart2014, esquivel2015} was measured with Alfv{\'e}n Mach number $M_A=V_L/V_A$, where $V_L$ is the injection velocity and $V_A$ is the Alfv{\'e}n velocity. Therefore $M_A\sim 1/B$ for the units adopted in this paper and the exact values can be found in table \ref{table:models}.

\begin{figure}[h]
\centering\includegraphics[width=\linewidth,clip=true]{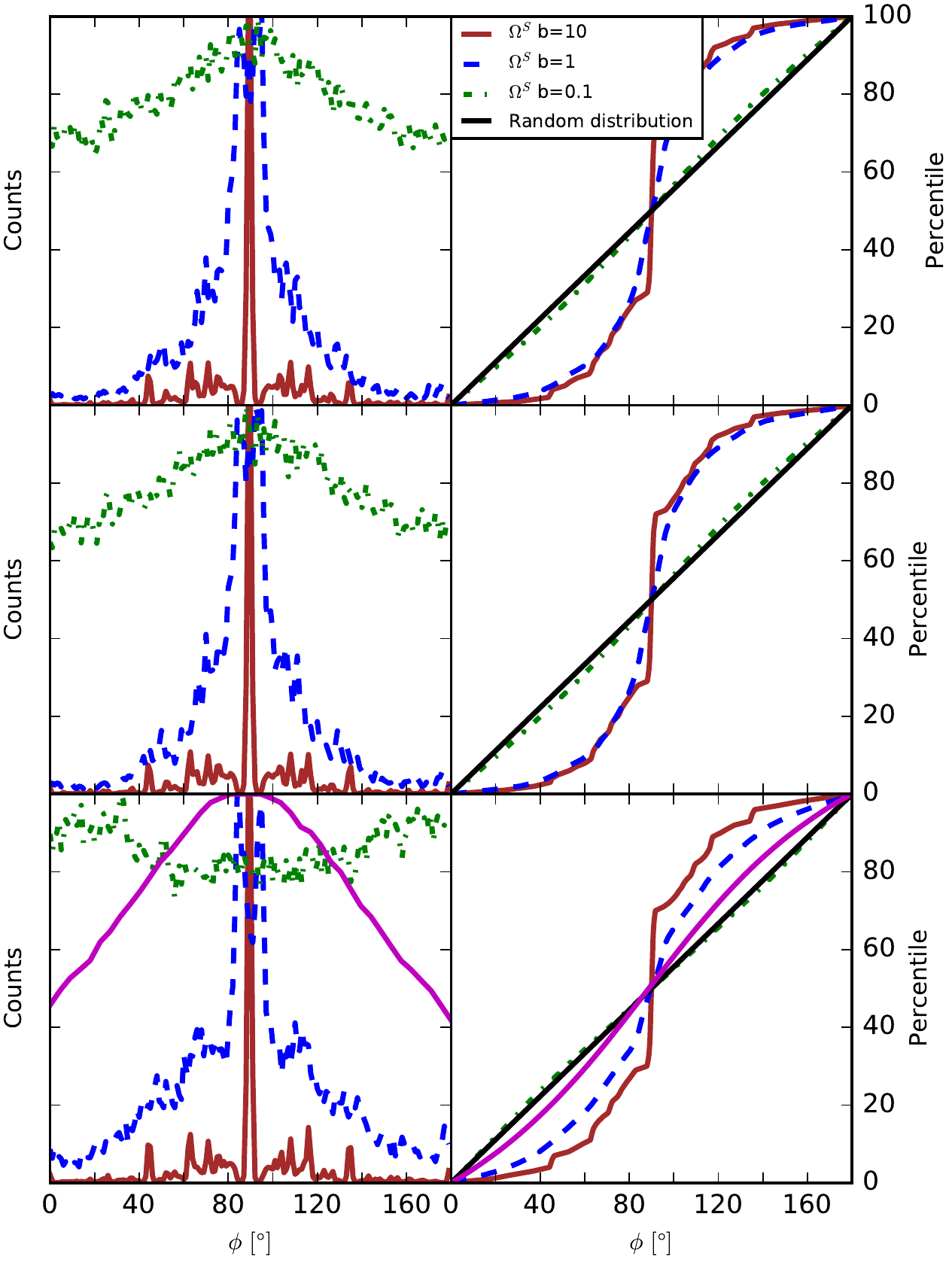}
\caption{Histograms ({\it $1^{st}$} column) and cumulative distributions ({\it $2^{nd}$} column) for the un-normalized velocity centroid ($S$;{\it top row}), the normalized velocity centroid ($C$; {\it middle row}), and the intensity ($I$; {\it bottom row}) using the three magnetization levels.  The distribution maps use all data points form the simulation. The histograms are normalized to 1, to account different data samples.  In {\it magenta} is the angle distribution of the density gradient technique developed by \citet{soler2013} corresponding to $M_A\approx3.1$ }
\label{fig:histogram}
\end{figure}

Figure \ref{fig:histogram} shows the cumulative distribution and the histogram for the angle measurements of the velocity centroids and intensity.  The histograms are normalized to have the maximum value set to one to better compare between different resolutions.  Table \ref{table:vgt-values} gives the standard deviation, $\sigma_\phi$, of the angle distribution for both $C$ and $S$.  Since $C$ and $S$ have similar angles distributions future analysis will center on the un-normalized velocity centroid (Section \ref{sec:properties}). Figure \ref{fig:histogram} quantitatively illustrates the correlation of the magnetic field with the velocity gradients measured by velocity centroids, as well as the variation of this correlation with the level of magnetization. 

The different panels of Figure \ref{fig:histogram} present the results using different styles. The first column shows the histogram or probability distribution that shows that for most cells the velocity gradients and the intensity tend to be aligned perpendicular to the magnetic field and that higher magnetization results in better alignment.  The second column present the same information in their cumulative distribution.  Cumulative distributions present information in percentiles rather than in counts.  The percentile indicates the value below which a given percentage of observations in a group of observations fall. The 50th percentile is the same as the median, at witch point half of the observations are below that observation \citep{zwillinger1999}. The black line corresponds to no correlation and the deviations from this line are proportional to the alignment of the magnetic field.  All panels show that the correlation for $B=0.1$ is marginal as expected for super-Alfv{\'e}nic turbulence for which the GS95 scalings are not applicable for large scales of motions.

\begin{table}
\centering
\caption{VGT parameters} 
\begin{tabular}{r*{9}{c}}
Model &  $\mathbf{B_o}$  & $\sigma^S_\phi$ & $\sigma^C_\phi$ & $\sigma^I_\phi$\\
\hline
\hline
1 & 10    & 24\degree & 23\degree & 27\degree\\
2 & 1      & 26\degree & 26\degree & 36\degree\\
3 & 0.1   & 49\degree & 49\degree & 53\degree\\
\hline
\end{tabular}
\label{table:vgt-values}
\end{table}

\subsection{Alignment of Density Gradients and Magnetic Fields} \label{sec:results2}

The velocity gradient technique traces the intrinsic velocity gradient present in a turbulent medium.  The turbulence also has its imprint on the density distribution \citep[see][]{bersnyak2005}.  A similar analysis to the one used on the velocity centroids is apply to the density (intensity), $I(\mathbf{x})$.   As shown in Figure \ref{fig:histogram}, the density gradient is not well correlated with the direction of the magnetic field, giving much larger error estimates for the direction of the magnetic field (Table \ref{table:vgt-values}).

The correlation of the density gradients and magnetic field was first noted by \citet{soler2013}. Their gradients were calculated using the square neighborhood around the cell, with a kernel of $3\times3$.  In Figure \ref{fig:histogram}, we present the correlation of column density gradients with the magnetic field.  Our earlier studies \citep[see][]{bersnyak2005,  kowal2007} suggest that velocity traces magnetic fields better and therefore velocity gradients should provide a more accurate direction of the magnetic field. 
Note that simulations in \citet{soler2013} included self gravity, hence one of the difference on the correlation of their and our density gradient. For the simulations without self-gravity the problematic nature of using density gradients is expected to increase for high Mach number turbulence when the density fluctuations lose clear correlation with magnetic field.  Low contrast density fluctuations according to \citep{beresnyak2005} follow GS95 picture, but large contrast fluctuations may be perpendicular or not well aligned. 

\section{Properties of the Velocity Gradient} \label{sec:properties}

\subsection{Reduction Factor}

In what follows, we introduce the reduction factor (RF) that measures the correspondence between two velocity gradients and magnetic fields.  This RF analogous to the Rayleigh reduction factor in dust alignment theory suggested by \citep{greenberg1968}.  Our RF is defined as:
\begin{equation}
R = 2\Big\langle cos(\phi)^2-\frac{1}{2} \Big\rangle \;,
\end{equation}
where $\phi$ is the angle between velocity gradients and magnetic field. The difference with the Reyleigh reduction factor is that we introduce our RF for the 2D distribution rather than for the 3D one. R gets 0 for no alignment and gets 1 when the gradients are perpendicular to the projected magnetic field.

MHD turbulence theory supports the presence of a velocity gradient perpendicular to the magnetic field. Since RF is a squared quantity of the cosine of the angle it does not distinguish the direction of the vectors (that form the angle), making it advantageous to characterize the accuracy of the VGT.  

\begin{figure}[h]
\centering\includegraphics[width=\linewidth,clip=true]{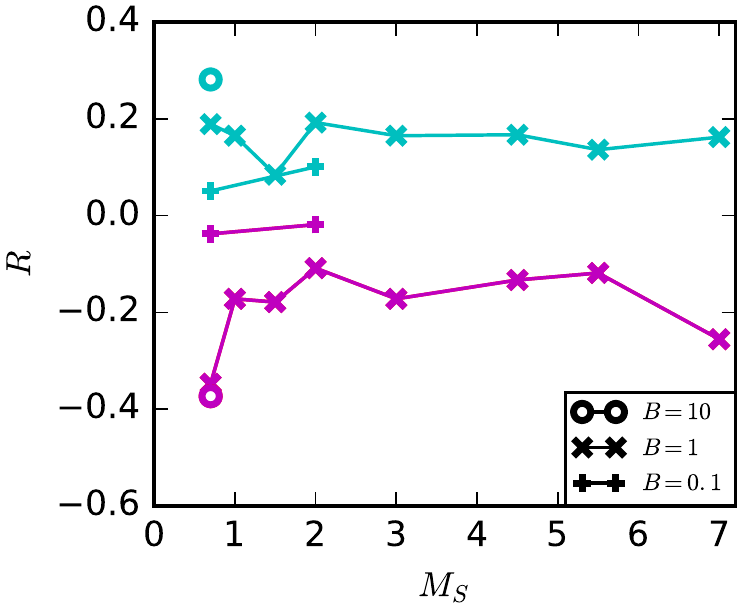}
\caption{The reduction factor, as a function of the sonic mach number ($M_S$) for the three different levels of magnetization.  In {\it cyan}: $R$ for the angle between $\mathbf{\Omega}^S$ and $\mathbf{\Omega}^I$ , in {\it magenta}: $R$ for the angle between $\mathbf{\Omega}^S$ and $\mathbf{B}$} 
\label{fig:reductionfactor}
\end{figure}

 We use the RF to measure the correspondence between the velocity centroid gradients, the intensity gradient and the magnetic field. The values of the RF for all models are presented in Figure \ref{fig:reductionfactor}. The negative value of $R$ for the velocity centroids and column density gradients mean that they tend to be aligned. The positive value of $R$ between the velocity centroid gradient and magnetic field mean that the velocity gradients tend to be perpendicular to magnetic field, as is expected in theory.

\subsection{Effects of the $M_S$ on the VGT}\label{sec:effectsofms}

The eddy velocity depends on the Mach number $M_S$, of the medium.  Higher $M_S$ is expected to weaken the correspondence between the velocity gradient and the magnetic field, as in the case of supersonic turbulence, the presence of shocks change the properties of the velocity gradient.  

\begin{figure}[h]
\centering\includegraphics[width=\linewidth,clip=true]{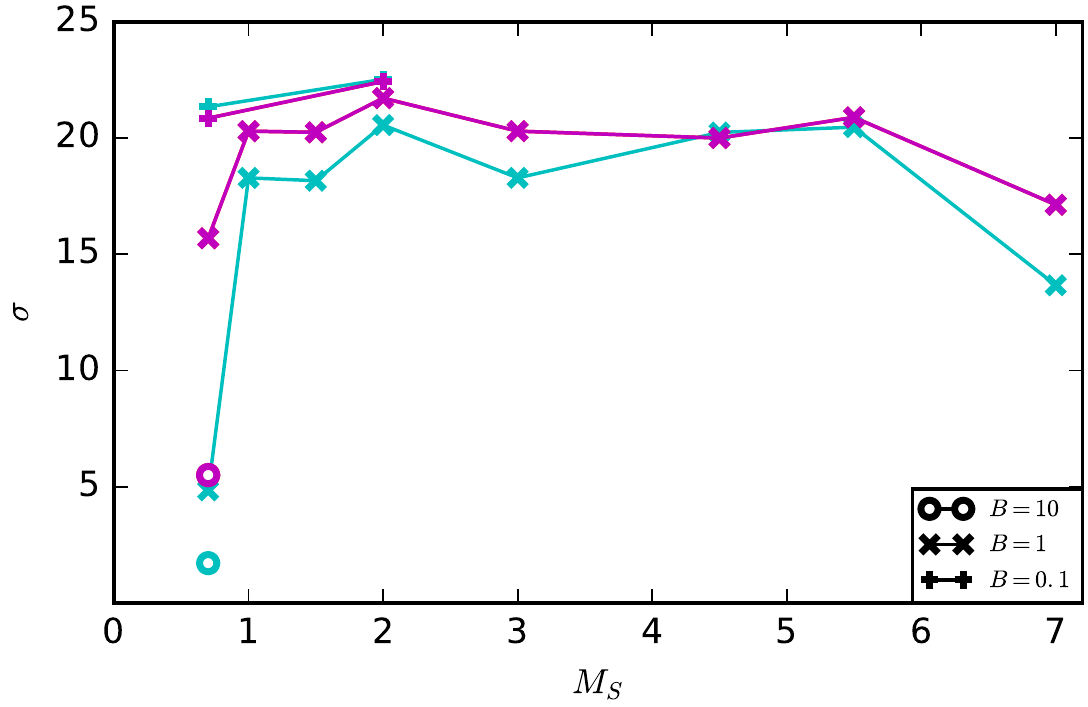}
\caption{The dispersion, measured with the standard deviation ($\sigma$), as a function of the sonic Mach number ($M_S$) for $\mathbf{\Omega}^S$ and $\mathbf{\Omega}^I$.  In {\it magenta}: $\sigma^I$, in {\it cyan}: $\sigma^S$.} 
\label{fig:ms-distribution}
\end{figure}

To analyze the effects of the $M_S$ on the velocity centroid we measure the spread of their distribution ($\sigma$).  The spread measurements only uses the velocity gradient measurements.  This spread, $\sigma$, most not be confuse with the spread of the angle distribution ($\sigma_\phi$) that requires both velocity gradient measurements and magnetic field measurements.  While both measurements are affected by $M_S$ and $M_A$, $\sigma$ is an intrinsic property of the velocity gradient.  With only the measurement of $\sigma$, both Mach numbers in the media are determined. This standard deviation we suggest to use to measure the strength of magnetic field (see section \ref{sec:c-f}).  Figure \ref{fig:ms-distribution} shows $\sigma$ as a function of $M_S$ for different $M_A$. We do not see the theoretically expected increase of the dispersion with $M_S$, which we attribute to the insufficient accuracy of our measurements. A more detailed study of the effect is going to be presented elsewhere.

\subsection{Interferometric Studies Using Velocity Gradients}

Gradients can be measured using interferometers and their advantage is that it is not necessary to restore the images first. For instance in \citet{gaensler2011} and \citet{burkhart2012} synchrotron polarization gradients were used for studying turbulence for the data with single dish observations missing. 

The velocity centroids can be calculated using raw interferometric data \citep[see][]{kandel2016a}. Therefore gradients can be obtained with this data and used to trace magnetic fields in distant objects, in other galaxies. Examples of this application will be presented in future work.

\subsubsection{The Effects of the Base Line in the Gradient Calculation}

Spectroscopic observations in its PPV format permits the measurement of velocity properties of the medium, allowing for measurements of the turbulent velocity field. This is fundamental since density measurements do not always trace the turbulent media and its properties \citep{chepurnov2008}.  To understand the limitations and scales needed in spectroscopic observations to use the VGT we transform the simulated data and remove the long and short base lines. This is done using a Fourier transformation on the PPV data cube and filtering different scales.  We remove information at 10\%, 5\% and 1\% of the inner cells (small $k$ long $l$) and outer cell (small $l$ long $k$).  Once the information is remove we reconstruct the plane of the sky for the velocity and the intensity, and then apply the VGT.  The removal of the long $k$ does not affect the technique giving the same value for the $\sigma$ for all cases. Therefore is important to have high resolution at small scales. 

\section{Obtaining Magnetic Field Intensity: Analog of Chandrasechar-Fermi technique}\label{sec:c-f}

The C-F technique and its later modifications have been used to determine the intensity of the magnetic field on the plane of the sky using the velocity dispersion and dust polarization measurements \citep[see][and ref therein]{falceta2008}.  Here we use the fact that velocity gradients similar to aligned grains tend to align perpendicular to the magnetic field. Therefore here we propose a new technique that is analogous to the C-F  but the gradients of velocity centroids are used instead of polarization vectors. The Equation \ref{eq:cf} can be rewritten in terms of the velocity centroid for the velocity dispersion, and the angle dispersion from the velocity gradient ($\sigma^U$):
\begin{equation}
B = \gamma \sqrt{4 \pi \rho} \frac{\delta v}{\sigma^U}  \;, 
\label{eq:cf2}
\end{equation}
where $\gamma$ is the correction factor that can potentially depend on $M_S$ and other parameters, e.g. the self-gravity effects.  In the classical C-F method there is also an analog of our $\gamma$. For instance, the factor $\sim 0.5$ was suggested using numerical simulations in \citep{ostriker2001}.  More accurate expressions have also been suggested \citep[see]{falceta2008}. However, for the sake of simplicity, in this study we use Eq. (\ref{eq:cf2}).

\begin{figure}[h]
\centering\includegraphics[width=1\linewidth,clip=true]{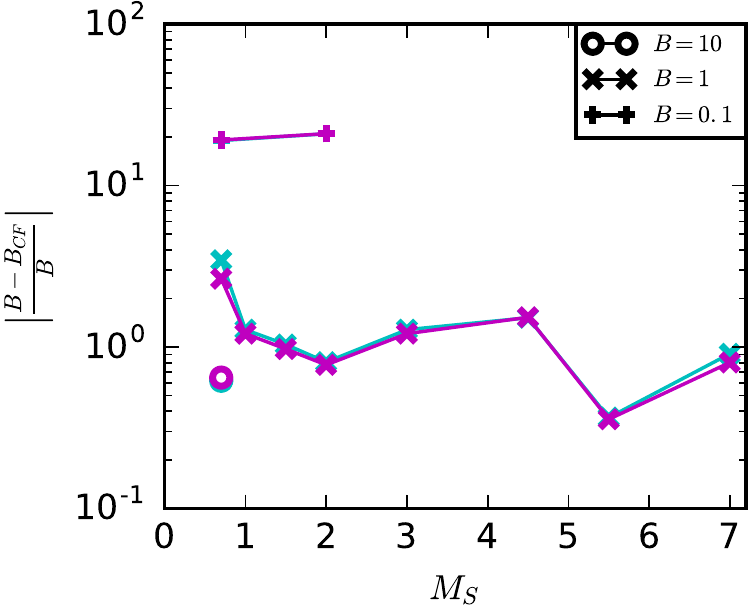}
\caption{The relative error estimation between the magnetic field (from the simulation) and the estimation from the C-F method as a function of $M_S$ for the different magnetization levels. In {\it magenta}: $\sigma^I$ and  in {\it cyan}: $\sigma^S$}
\label{fig:c-f}
\end{figure}

In order to determine the $\gamma$ parameter for our technique we calculate the values for Equation \ref{eq:cf2} for the first three models (same $M_S$ but different $M_A$). $\delta v$ is the dispersion of the velocity field measured by the normalized velocity centroid and $\sigma^U$ is the dispersion of the velocity gradient (see Figure \ref{fig:ms-distribution} for the different values). $\gamma$ for the intensity ($I$), and both velocity centroids is $\sim1.29$.  The reason behind a larger parameter for the modified C-F method, is a larger dispersion of the velocity gradient than the dispersion found in dust alignment.  Figure \ref{fig:c-f} shows the relative error estimation on the projected magnetic field.  The value of $\gamma$ is an average of the values obtain for the different gradients. Our results should be treated only as a demonstration of the potential applicability of the technique. For instance, in Figure \ref{fig:c-f} the errors for supersonic turbulence seem smaller than those of subsonic turbulence, this is a pure effect of the $\gamma$ choice and not a condition on the turbulence properties. We expect that the errors of the technique can be significantly reduced by using a more sophisticated expressions for magnetic field strength.

To account for different degrees of compressibility in the medium we use all data cubes (Table \ref{table:models}).  We know that the velocity gradients dispersion is highly correlated with the $M_A$ and loosely with the $M_S$ (Section \ref{sec:effectsofms}), hence some dependance is expected. Since the C-F method also incorporates the spread of the velocity field, the overall measurements depend weakly on $M_S$ not requiring extra parametrization. 

\section{Additional effects} \label{sec:discussion}

Further analysis to the properties to the velocity gradient are done with the correlation functions and statistical moments (see appendix \ref{appen:extra}).  They all show that the velocity gradient is susceptible to different degrees of magnetization and that the velocity gradient aligns perpendicularly to the magnetic field. 

\subsection{Telescope resolution effects}

Finite telescope resolution introduces additional uncertainties. In order to account for data averaging within the telescope beam, we use two different smoothing kernels on the velocity centroids -- a square and a Gaussian kernel. For the square kernel, each point in the velocity centroid is replaced with the average of the points in its vicinity.  In this case, the vicinity was defined as square boxes of lag $r= 2^n$ with $n$ from 0 to 6.  The velocity gradient is then estimated using the smooth velocity centroids and the un-smoothed magnetic field.  This process reduces or increases the different values as seen in Table \ref{table:smooth}.  The process of smoothing is thus not a technique to enhance the results in all cases. 

\begin{table}[]
\centering
\caption{Smoothing effects of the square kernel} 
\label{table:smooth}
\begin{tabular}{ll|l}
                                                                       & \multicolumn{1}{c|}{\begin{tabular}[c]{@{}c@{}}Model\\ B = 10\end{tabular}} & \multicolumn{1}{c}{\begin{tabular}[c]{@{}c@{}}Model \\ B = 1\end{tabular}} \\  \hline
\multicolumn{1}{l|}{\begin{tabular}[c]{@{}l@{}}lag\\ `r'\end{tabular}} & $\sigma^S$ & $\sigma^S$                 \\ \hline \hline
\multicolumn{1}{r|}{0}                         & 24\degree                & 39\degree                                       \\
\multicolumn{1}{r|}{2}                         & 26\degree                & 39\degree                                        \\
\multicolumn{1}{r|}{4}                         & 30\degree                & 40\degree                                       \\
\multicolumn{1}{r|}{8}                         & 34\degree                & 41\degree                                        \\
\multicolumn{1}{r|}{16}                       & 35\degree                & 43\degree                                        \\
\multicolumn{1}{r|}{32}                       & 33\degree                & 47\degree                                        \\
\multicolumn{1}{r|}{64}                       & 32\degree                & 51\degree                                         \\ \hline                              
\end{tabular}
\end{table}

\begin{table}[]
\centering
\caption{Smoothing effects of the Gaussian kernel} 
\label{table:gauss}
\begin{tabular}{rl|l}
\multicolumn{1}{l}{}      & \multicolumn{1}{c|}{\begin{tabular}[c]{@{}c@{}}Model\\ B = 10\end{tabular}} & \multicolumn{1}{c}{\begin{tabular}[c]{@{}c@{}}Model \\ B = 1\end{tabular}} \\ \hline 
\multicolumn{1}{l|}{FWHM} & $\sigma^S$& $\sigma^S$                               \\ \hline \hline
\multicolumn{1}{r|}{0}    & 24\degree                                    & 39\degree                          \\ 
\multicolumn{1}{r|}{2}    & 29\degree                                    & 40\degree                           \\
\multicolumn{1}{r|}{4}    & 35\degree                                    & 41\degree                           \\
\multicolumn{1}{r|}{8}    & 40\degree                                    & 43\degree                           \\
\multicolumn{1}{r|}{16}  & 41\degree                                    & 47\degree                           \\ \hline
\end{tabular}
\end{table}

The Gaussian smoothing kernel was used to simulate more realistic observational data.  Namely, observational data does not have pencil-thin beam resolution, but more of a smooth beam resolution.  We use 4 values for the full-width-half-maximum (FWHM) of 2, 4, 8 and 16 (see Table \ref{table:gauss}).  

\subsection{Noise}

Realistic observational data has intrinsic noise. We add artificial noise to the simulations to understand the changes on the VGT.  We introduce Poisson noise to the projected data (the plane of the sky) with signal-to-noise (S/N) ratios of 10, 50, 100 and 400.  The noise is added independent to both the intensity and velocity maps.  Using the same procedure described in subsection \ref{subsec:vgt}, we obtain the gradient for the velocity centroid and the intensity for all five models.  We then estimate the standard deviation for the different noise levels.  For most cases the changes in the standard deviation are less than 0.5\degree compare to the noise-less data. Table \ref{table:noise} shows the change in the standard deviation between the case with a S/N of 10 to the noise-less data.

\begin{table}[]
\centering
\caption{Effect of noise in the $\sigma$ estimation}
\label{table:noise}
\begin{tabular}{cc|c}
        & \multicolumn{1}{c|}{$I$}                      & \multicolumn{1}{c}{$S$}  \\
        & $|\Delta \sigma_\phi|$ & $|\Delta \sigma_\phi|$ \\ \hline \hline
Model 1 & 0.15\degree   & 0.14\degree     \\ \hline
Model 2 & 0.08\degree   & 1.35\degree     \\ \hline
Model 3 & 0.18\degree   & 0.66\degree     \\ \hline
Model 4 & 0.38\degree   & 1.05\degree     \\ \hline
Model 5 & 0.34\degree   & 0.34\degree     \\ \hline
\end{tabular}
\end{table}

\subsection{Column Density Effects}

\begin{figure}[h]
\centering\includegraphics[width=\linewidth,clip=true]{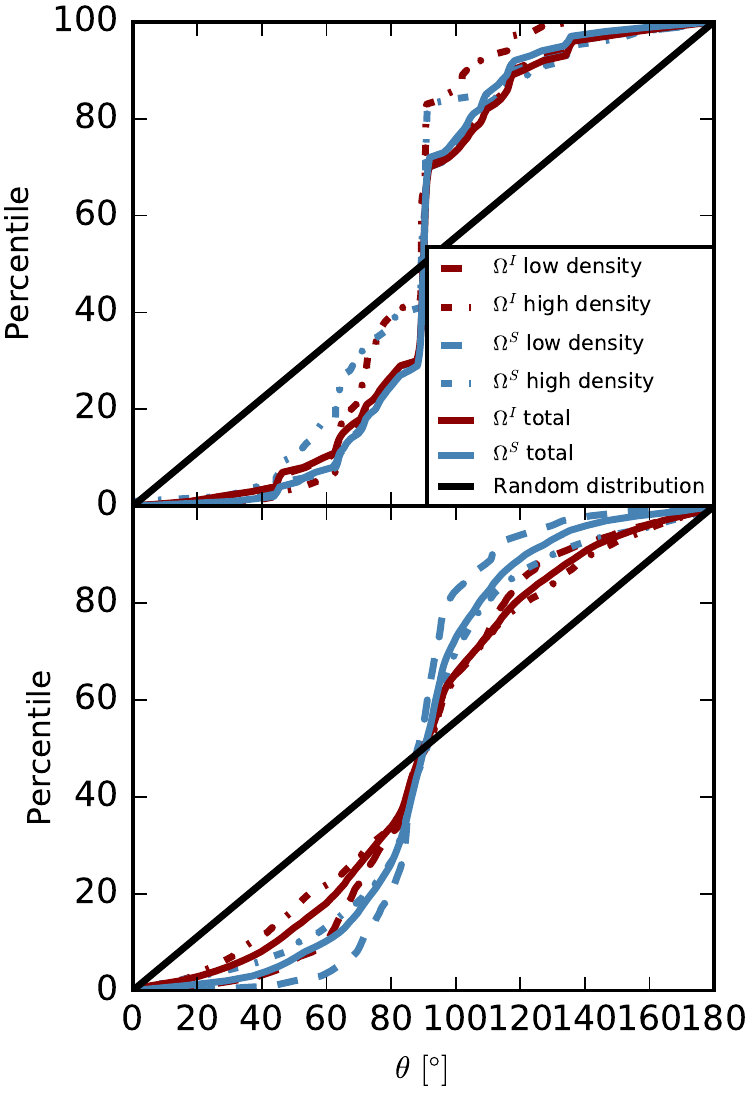}
\caption{Cumulative pseudo-angle distribution for the case of $B=10$ {\it top panel} and $B=1$ {\it bottom panel}.  In {\it dark red} the gradient for the intensity and in {\it blue} the gradient for the un-normalized velocity centroid for all densities. Straight lines correspond to the gradient for all column density, dash lines for low column density and dot-dash for high column density.} 
\label{fig:colum-dens}
\end{figure}

Measurements of magnetic fields strengths are affected by column density effects  \citep[][]{evans1999, clark2014, ntormousi2016, planck2016}.   To understand the effects of column density in the VGT we divided the data sample in low-, mid- and high- column density for models 1 and 2 (Table \ref{table:models}).  The gradient for the intensity and the un-normalized velocity centroid as calculated in Section \ref{sec:results1}, are distributed into three groups using the column density criterion.  Using only the low and high column density data, the cumulative distribution is obtained as shown in Figure \ref{fig:colum-dens}. For both degrees of magnetization and both gradients, $\mathbf{\Omega}^I$ and $\mathbf{\Omega}^S$, the low column density has smallest errors, followed by the global calculation, the high column density (Table \ref{table:column}).  It is important to note that this simulations are done in the subsonic regime where there are no shocks in the medium. In the presence of shocks, the velocity gradient can align with the magnetic field, changing the properties of the technique (see Subsection \ref{sec:effectsofms}).

\begin{table}[]
\centering
\caption{Column density effects}
\label{table:column}
\label{my-label}
\begin{tabular}{lll|ll}
                          & \multicolumn{2}{c|}{$\sigma^ S$} & \multicolumn{2}{c}{$\sigma^ I$} \\
                                               & low              & high                  & low                   & high                  \\ \hline
\multicolumn{1}{l||}{B=10} & 25\degree    & 25\degree          & 20\degree         & 21\degree                     \\
\multicolumn{1}{l||}{B=1}   & 20\degree    & 21\degree          & 32\degree         & 38\degree                 
\end{tabular}
\end{table}

\subsection{Sub-Block Analysis}

\begin{figure}[h]
\centering\includegraphics[width=\linewidth,clip=true]{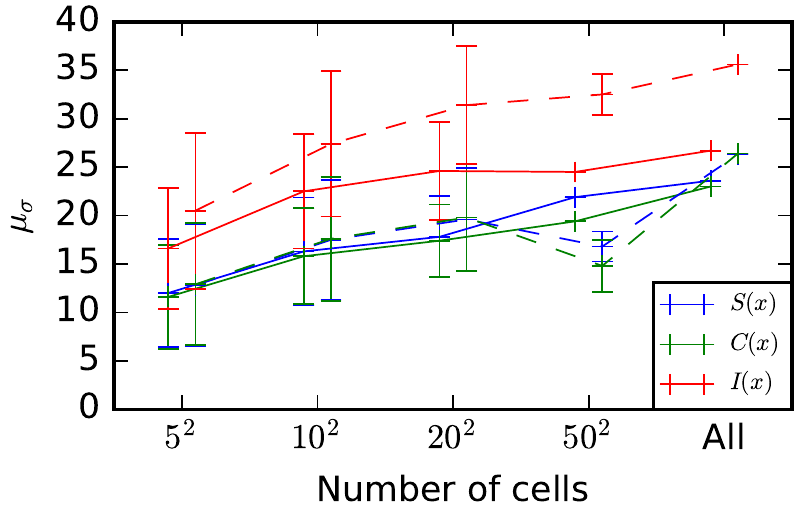}
\caption{The mean ($\mu_\sigma$) of the $\sigma$ for de velocity and intensity gradients, for each of the different sub-blocks as a function of the number of cells in the sub-block.  $\sigma$ is the spread of the velocity gradient in each sub-block.  The error bars correspond to the variations in the values of the of the $\sigma$ ($\sigma_\sigma$).  In dash lines the B=1 case and solid lines the B=10 case.  The points on the plot are displaced from their respective number of cell, to allow a better visualization of the values.} 
\label{fig:subblock}
\end{figure}

The velocity gradient technique can asses the direction of the magnetic field with good point wise correspondence.  The values previously reported use the full simulated domain, but it is important to understand the limitations of smaller data samples.  To analyze this effect we subdivided the simulation into individual regions with different numbers of cells. At each region, we apply the VGT and obtain the velocity gradient.  With the velocity gradient we obtain the standard deviation of the angle distribution ($\sigma_\phi$).  For most regions the standard deviation, $\sigma_\phi$, was smaller than the one found in the full simulated box.  To understand how the different standard deviation vary as a function of the number of cells, we estimate the spread ($\sigma_{\sigma_\phi}$) and the mean ($\mu_{\sigma_\phi}$) of the different $\sigma_\phi$ along the different regions.  Figure \ref{fig:subblock} shows the mean, $\mu_{\sigma_\phi}$,  as a function of the number of cells, and the error bars correspond to the standard deviation, $\sigma_{\sigma_\phi}$.  It is clear that our technique is effective even with small data samples the technique works.  If  the observed or simulated media have shocks, a sub grid analysis that removes (or properly analyzes) these regions would give a better overall performance of our technique.

\section{Comparison with Other Techniques}

Many other techniques to study magnetic fields have limitations.  The measurements of polarization arising from aligned dust  are done in the optical/near-IR for stars and far-IR/sub-mm for dust.  The interpretation of these measurements requires an understanding of the dust alignment and modeling its failure for high optical depths.  In spite of the significant progress of grain alignment theory \citep[see][]{lazarian2007, lazarian2008, hoang2008, hoang2009, hoang2016} the magnetic field studies employing dust polarimetry are not straightforward. In some cases it is necessary to provide the detailed modeling of the radiation field as if the radiative field is insufficient, the grain alignment fails and does not reveal magnetic fields \citep[see reviews by][]{andersson2015, lazarian2015}. In addition, high resolution, high sensitivity polarization data is not readily available. For instance, a great dust polarization map has been obtained by \citet[][see references therein]{plank2016} To get a map with better resolution one needs to wait years for the next analogous mission. Tracing magnetic fields with higher resolution using the VGTI may be much easier.

In addition, the classical C-F method requires both polarization and spectroscopic velocity measurements to determine the intensity of the magnetic field. The technique that we suggested using velocity centroid gradients only requires the velocity measurements, which is a simplification.

The direction of magnetic fields can also be obtained using statistical techniques that use the predicted anisotropies of MHD turbulence \citep{lazarian2002, esquivel2005}. These variation of this techniques using the Principal Component Analysis (PCA) was used and show that the magnetic fields are in agreement with the measurements using dust polarization \citep{heyer2008}. However, being statistical in nature, the techniques are able to provide average magnetic-field directions. In comparison, the technique of tracing magnetic fields with the velocity gradients that we introduced in this page provides a more detailed information of the magnetic field. The synergy of all these techniques is to be revealed in the future publications.

We would like to stress the exploratory nature of our present study. We did not seek to provide detailed prescriptions for better tracing of magnetic fields and for obtaining its intensity from spectroscopic observations. This is the goal of further studies. Instead, we introduced a new way of tracing magnetic fields and showed its practical applicability using synthetic observations. This way we obtained encouraging results which stimulate further in depth studies.

We also would like to emphasize that velocity gradients should not be treated just as proxies of the magnetic field direction or just an alternative technique for tracing magnetic field without polarimetric measurements. Velocity gradients are  the measures of the interstellar physics. For instance, they are expected to respond to shocks and self gravity in different way compared to magnetic fields. Therefore the misalignment of velocity centroids gradients and dust polarization may be very informative. Similarly, the studies of relative alignment of the velocity centroid gradients and the column density gradients opens a new avenue for exploring interstellar physics. Thus we expect the three measures: velocity gradients, density gradients and dust polarization to be used simultaneously whenever possible.

The synergy of the VGT and other techniques giving magnetic field direction is still to be explored. Some advantages are obvious even now. For instance, the Goldreich-Kylafis technique as well as the technique based on atomic/ionic alignment \citep[see][]{yan2015} can provide magnetic field with the ambiguity of 90 degrees, which may be confusing. The VGT may be used to remove the ambiguity. We do not expect the qualitative nature of the VGT technique to be changed in the presence of self-absorption \citep[see][for more details]{lazarian2006}. In the conditions where the infall induced by self-gravity is important, the alignment of velocity gradients and magnetic field can change. This issue is studied elsewhere.  

Our work shows the advantages of using theory motivated approach for developing techniques for studying magnetic fields and turbulence from observations. The prediction that velocity gradients are expected to be perpendicular to magnetic field follows from the MHD turbulence theory provided that the Alfv{\'e}nic modes dominate the contribution of the fast modes \citep[see][]{cho2002, cho2003}.  In this paper we proved that this provides a new way to study magnetic fields observationally.

In general, we believe that the synergy of different techniques is the best for tracing magnetic fields and our new techniques, namely, the VGT, can be useful for studying magnetic fields in the diffuse ISM and molecular clouds.

\section{Conclusions} \label{sec:conclusion}

This work presents a new technique, the VGT, of tracing magnetic field and of estimating its magnitude using only spectroscopic data.  The method is based on the fact that the eddies align with the local 3D magnetic field, and that this creates eddies velocity gradients perpendicular to the direction of the field. To test the technique we use synthetic observations constructed with 3D MHD simulations. Further analysis with a larger set of initial conditions should be explored to fully understand the implications and limitations of this technique.  A summary of the work can be given as follows:

1.- For observational studies, the velocity gradients can be represented by the velocity centroids gradients $\mathbf{\Omega}$ that we shown to trace reliably  the direction of the projected magnetic field in a sub- and trans-Alfv{\'e}nic regime.

2.- We proposed and successfully tested  a new technique of estimating the level of magnetization of the media given by the Alfv{\'e}n Mach number $M_A$ and the magnetic field intensity. This new technique only requires spectroscopic velocity data and does not require any polarimetry measurement. 

3.-  We showed that the VTG can work in the presence of averaging arising from finite telescope resolution. The VGT can also employ interferometry data with some of the baselines missing. This opens the prospects of using the VGT for a wide variety of objects.

4. Our work suggest the synergy of the simultaneous use of the VGT, polarimetry data, density gradients simultaneously as their use in combination with other techniques to study magnetic fields in order to explore the turbulent magnetized interstellar medium.

\acknowledgments 
We thank Jungyeon Cho for many insightful discussions related to our numerical results. We also thank the discussions with Blakesley Burkhart, Julie Davis, Dhanesh Krishnarao and the referee, and data cubes sent to us by Grzegorz Kowal. AL and DFGC are supported by the NSF grant AST 1212096. Partial support for DFGC was provided by CONACyT (Mexico). AL acknowledges a distinguished visitor PVE/CAPES appointment at the Physics Graduate Program of the Federal University of Rio Grande do Norte and thanks the INCT INEspa\c{c}o and Physics Graduate Program/UFRN, at Natal, for hospitality. 


\begin{appendices}
\section{Extra properties of the velocity gradients}\label{appen:extra}

The velocity gradients are affected by noise in the data, telescope beam effects and column density effects, that change their observational properties, but they also present effects related to the level of the magnetization. Here we detail how the properties of the velocity gradient change.  The change is measured with correlation functions and statistical moments.
\subsection{Statistical moments of the velocity gradient}\label{appen:moments}

The different levels of magnetization produced in the medium modify the distribution of the velocity gradient, $\mathbf{\Omega}$.  These differences on the distribution are quantized by the statistical moments.  We use the L-moments to understand its properties and relate them to the Alfv{\'e}nic mach number. L-moments, introduced by \citet{hosking1990}, measure the properties of the distribution like regular statistical moments.  The L-moments use are as defined in \citep{wang1996} and the L-moment ratios are $t_3 = l_3/l_2$ and  $t_4 = l_4/l_2$:

\begin{eqnarray*}
l_1 =                    \frac{1}{C^n_1}\sum^n_{i=1}x_i\;, \hfill \nonumber \\
l_2 = \frac{1}{2}\frac{1}{C^n_2}\sum^n_{i=1}\Big(C^{i-1}_1-C^{n-1}_1\Big)x_i\;, \hfill \nonumber \\
l_3 = \frac{1}{3}\frac{1}{C^n_3}\sum^n_{i=1}\Big(C^{i-1}_2-2C^{i-1}_1C^{n-1}_1-C^{n-1}_2\Big)x_i\;, \hfill \nonumber \\
{\scriptstyle l_4 = \frac{1}{4}\frac{1}{C^n_4}\sum^n_{i=1}\Big(C^{i-1}_3-3C^{i-1}_2C^{n-1}_1+3C^{i-1}_1C^{n-1}_2-C^{n-1}_3\Big)x_i}\;, \hfill  \\
\label{eq:lmom}
\end{eqnarray*}
where $x_i$ is the data sample, and $C^n_m$ is the number of combination of $m$ items from $n$ defined as:
\begin{equation} 
C^m_k = \frac{m!}{k!(m-k)!}\;. \hfill
\end{equation}
Because L-moments are a linear function of the data, they are less susceptible to sampling variability (such as outliers in the data) than conventional moments.  The L-moment ratios such as L-skew ($t_3$) and L-kurtosis ($t_4$) have the property $|t_i|<1$.

\begin{figure}[h]
\centering\includegraphics[width=\linewidth,clip=true]{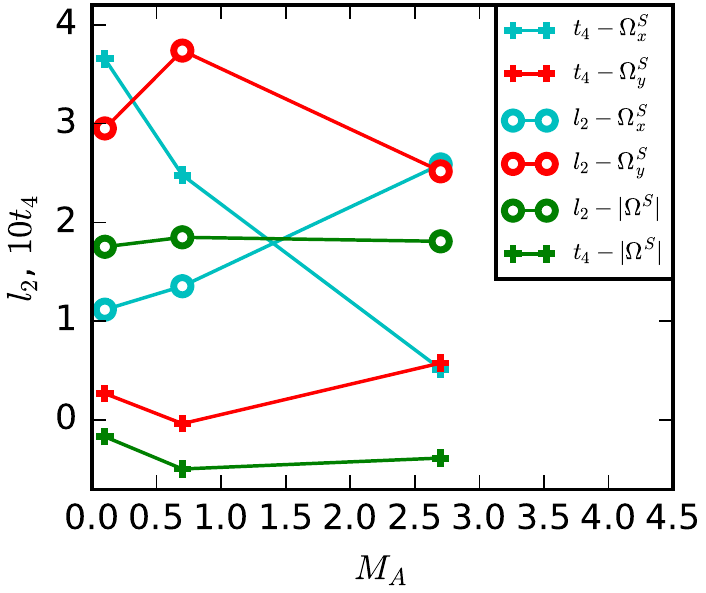}
\caption{L-kurtosis ({\it circles}) and L-scale ({\it crosses}) for $\mathbf{\Omega}_x^S$ ({\it cyan}),  $\mathbf{\Omega}_y^S$ ({\it red}) , and $|\mathbf{\Omega}^S|$ ({\it green}) as a function of the Alfv{\'e}nic mach number. The L-moments for $\mathbf{\Omega}_x^S$ and $\mathbf{\Omega}_x^C$ have the same trends.}
\label{fig:lmoments}
\end{figure}

The two components of the velocity gradient (parallel and perpendicular to the mean magnetic field) and the magnitude of the gradient are analyzed separately using the L-moments for the full angle span (0\degree to 180\degree).  $\mathbf{\Omega}_x^S$ is the component of the velocity gradient that is parallel to the mean magnetic field while $\mathbf{\Omega}_y^S$ is in the perpendicular direction.  L-skew and L-mean are close to zero for all the cases of the velocity gradient components, giving no information on the intensity of the magnetic field.  The L-mean and L-skew for $\mathbf{\Omega}_x^S$ is always constant (as a function of $M_A$) since its distribution always has a peak around 90\degree (perpendicular to the magnetic field), for $\mathbf{\Omega}_y^S$ the lack of changes are due to no preferential direction of motion set by the magnetic field and therefore a more homogenous medium.  The L-kurtosis and L-mean are shown as a function of the Alfv{\'e}nic mach number in Figure \ref{fig:lmoments}.  The moments are only a function of $M_A$ for $\mathbf{\Omega}_x^S$ since it is the component susceptible to the changes in the intensity of the field.  The decrease in the L-kurtosis and increase in the L-scale for $\mathbf{\Omega}_x^S$ reflects that $\mathbf{\Omega}^S$ is mostly perpendicular to the magnetic field with most of its components around zero. This implies that given a distribution of the velocity gradient, measuring the L-moments for both components one can estimate the intensity of the magnetic field and its global direction.  To estimate the mean field direction it is necessary that the components of the velocity gradient match those presented here.

\subsection{Anisotropy}\label{appen:anis}

Correlation functions are a two-point statistical tool.  In a turbulent medium they can be used to measure the power spectrum of the energy cascade, and to analyze the anisotropy of the medium \citep{esquivel2003, esquivel2005}:

\begin{equation} 
 CF(\mathbf{r}) = \bigg\langle f(\mathbf{x}) \cdot f(\mathbf{x + r}) \bigg\rangle \;,
 \label{eq:corr}
\end{equation}
where $CF$ is the correlation function, $\mathbf{r}$ is the ``lag'', $\langle \cdot \rangle$ denotes the average over all points, and $f$ denotes the desired function -- in this case the velocity centroids, intensity and velocity gradient.  The power and energy spectrum can be estimated by a correlation function of the velocity centroid. 

\begin{figure}[h]
\centering\includegraphics[width=\linewidth,clip=true]{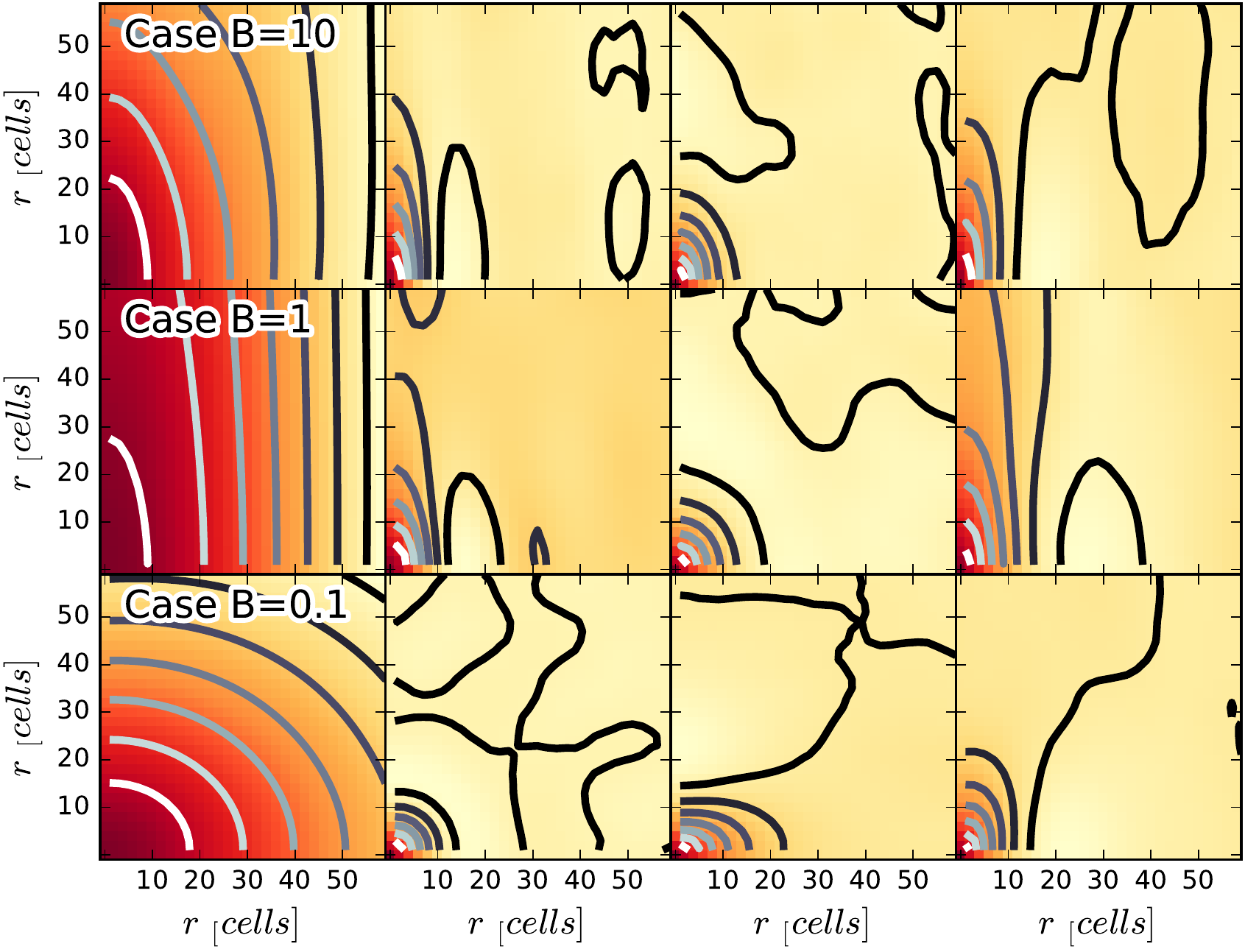}
\caption{Correlation function as a function of angle with respect to the mean magnetic field for all cases.  From left to right: $S$, $|\mathbf{\Omega}^S|$, $\mathbf{\Omega}_x^S$, and $\mathbf{\Omega}_y^S$.  From top to bottom the mean magnetic field intensities: 10, 1, and 0.1.} 
\label{fig:correlation}
\end{figure}

The anisotropies of the medium are measured using the correlation functions by making the lag a function of the angle to the global mean magnetic field, $\mathbf{r}(\theta)$. The angle $\theta$ has a span of 90\degree, going from $\mathbf{r}(0^{\circ}) = \mathbf{r}_\parallel$ ($\mathbf{r}_\parallel \times \mathbf{B} = 0$) to $\mathbf{r}(90^{\circ}) = \mathbf{r}_\perp$  ($\mathbf{r}_\perp \cdot \mathbf{B} = 0$). In an isotropic medium the values of the correlation function should be independent of the direction of the lag, i.e. $CF(\mathbf{r}_\parallel) = CF(\mathbf{r}_\perp)$.  In the case of an anisotropic medium, such as in Alfv{\'e}nic turbulence, the correlation presents a preferential direction.  The preferential direction is set by the mean magnetic field -- in other words, the correlation function changes depending on the direction of the mean magnetic field.   The intensity of the magnetic field determines the level of the anisotropy and hence the elongation of the isocontours of the velocity centroids (Figure \ref{fig:correlation}).  The mean direction of the magnetic field sets the elongation direction of the isocontours.  Even if the CF can trace the mean magnetic field, as several other techniques, is important to understand new ways to trace them as the VGT.  More over the CF require a lot of sample data to estimate roughly the direction, and so is important to find techniques that can map the magnetic field them in a smaller scale.

Applying the same method of correlation functions to the velocity gradient $\mathbf{\Omega}$, one can see that just as the velocity centroid is affected by the intensity of the magnetic field, so is the velocity gradient. Hence, the correlation function of the velocity gradient can determine the level of magnetization of the medium and the direction of the mean field to determine if the velocity gradient technique can be used.

\end{appendices}


\bibliographystyle{apj}
\bibliography{biblio}{}

\end{document}